# Triplet-Singlet Spin Relaxation via Nuclei in a Double Quantum Dot

A. C. Johnson[1], J. R. Petta[1], J. M. Taylor[1], A. Yacoby[1,2], M. D. Lukin[1], C. M. Marcus[1], M. P. Hanson[3], A. C. Gossard[3]

[1] Department of Physics, Harvard University, Cambridge, Massachusetts 02138, USA

[2] Department of Condensed Matter Physics, Weizmann Institute of Science, Rehovot 76100, Israel

[3] Materials Department, University of California, Santa Barbara, California 93106, USA

**The spin of a confined electron, when oriented originally in some direction, will lose memory of that orientation after some time. Physical mechanisms leading to this relaxation of spin memory typically involve either coupling of the electron spin to its orbital motion or to nuclear spins[1-7]. Relaxation of confined electron spin has been previously measured only for Zeeman or exchange split spin states, where spin-orbit effects dominate relaxation[8-10], while spin flips due to nuclei have been observed in optical spectroscopy studies[11]. Using an isolated GaAs double quantum dot defined by electrostatic gates and direct time domain measurements, we investigate in detail spin relaxation for arbitrary splitting of spin states. Results demonstrate that electron spin flips are dominated by nuclear interactions and are slowed by several orders of magnitude when a magnetic field of a few millitesla is applied. These results have significant implications for spin-based information processing[12].**

The coupling of nuclear spins to electrons in low-dimensional semiconductors is known from optical and transport studies in quantum Hall systems to yield rich physical effects and provide new probes of the relatively isolated quantum system of nuclear spins in solids[13-16]. Confined electrons interacting with relatively few nuclei are particularly sensitive to hyperfine coupling. This can lead to dramatic effects such as tunnelling currents that slowly oscillate in time and electrical control and readout of nuclear polarization[17,18]. Here, we show that the interaction between single electrons confined in quantum dots with ensembles of lattice nuclei can dominate electron spin relaxation.

We use high-frequency pulsed gates to measure spin relaxation in a GaAs double quantum dot (Fig. 1a). Measurements are performed near the (1,1) to (0,2) charge transition, where (n,m) denotes the absolute number of electrons on the left and right dots. In the (0,2) configuration, the two electrons form a spin singlet to avoid the large Pauli exclusion energy cost (0.4 meV >> kT ~ 10 μeV) of occupying an excited orbital state[19,20]. In the separated (1,1) configuration, the two electrons may occupy any spin state. That is, apart from any Zeeman energy (~2.5 μeV at 100 mT), the singlet, (1,1)S, and three triplets, (1,1)T_, (1,1)T$_0$, and (1,1)T$_+$ ($m_s$ = -1,0,1 respectively) are effectively degenerate given the weak interdot coupling to which the system is tuned.

Spin relaxation is measured by preparing an unpolarized mixture of (1,1) states and monitoring the probability of transition to (0,2)S after the latter is made lower in energy by changing the electrostatic gate configuration. The different local environments acting on the two spins cause the two-electron spin state to evolve in time, and only if this spin state passes near (1,1)S is a transition to (0,2)S allowed. The average occupancy of the left dot, which reflects the probability of this transition, is monitored using local quantum point contact (QPC) charge sensors[19]. Conductances $g_{ls}$ and $g_{rs}$ of the left and right sensors change by several percent when an electron enters the dot nearest the sensor[21-24].

Energy levels of each dot were controlled by voltage pulses on gates L and R, as shown in Fig. 1c[19]. The double dot was cycled through three configurations, depicted in Fig. 1e, while measuring the average QPC conductances. In the Empty ("E") step, the second electron is removed, leaving a (0,1) state. In the Reset ("R") step, a new second electron is added, initializing the (1,1) state to an unbiased mixture of the singlet, (1,1)S, and the three triplets $(1,1)T_-$, $(1,1)T_0$, and $(1,1)T_+$. In the Measurement ("M") step, (0,2) is lowered relative to (1,1) until (0,2)S becomes the ground state, while the (0,2) triplets remain inaccessible, above the (1,1) states. Because tunnelling preserves spin, only (1,1)S can relax to (0,2)S, while the (1,1) triplets are spin-blockaded from making this transition[25,26].

The measurement step accounted for 80% of the pulse period (E and R were each 10%) so the time-averaged charge-sensor signal mainly reflects the charge state during the measurement time, $t_M$. Figure 1d shows $g_{rs}$ as a function of the dc gate voltages $V_L$ and $V_R$ with pulses applied. The dashed lines indicate locations of ground-state transitions during the M step, as seen in unperturbed double dots[22]. Gate pulses alter this signal only within the "pulse triangle" (outlined by solid white lines). Here $g_{rs}$ is intermediate between the (0,2) and (1,1) plateaus, indicating that although (0,2) is the ground state, the system is often stuck in the excited (1,1) state. In the regions labelled M′ and M″, alternate, spin-independent relaxation pathways, shown in Fig. 1f, circumvent the spin blockade.

The magnetic field, $B$, and $t_M$ dependence of the charge sensor signal is shown in Fig. 2. With $t_M = 8$ μs, a large signal is seen in the pulse triangle indicating that some of the (1,1) to (0,2) transitions are spin blocked. As $t_M$ is increased this signal decreases (Fig. 2b), indicating that $t_M$ is approaching the (1,1) singlet-triplet relaxation time. This is accompanied by a reduction in the pulse triangle size due to thermally activated processes as in Fig. 1f. Similar data, but at $B = 0$, are plotted in Figs. 2c,d. The signal in the pulse triangle is noticeably weaker for the same $t_M$, particularly near the (1,1) to (0,2) charge transition, indicating enhanced spin relaxation.

Detailed measurements of residual (1,1) occupation as a function of detuning (the energy difference between the (1,1) and (0,2) states) are shown in Fig. 3. Conductances $g_{ls}$ and $g_{rs}$ were measured along the diagonal white line in the upper panel of Fig. 3, for various $B$ and $t_M$, and converted to occupation, <N>, by scaling to the average (1,1) and (0,2) levels outside the pulse triangle. Data are shown in detail for the points labelled A through D. As in Fig. 2, strong field dependence was found at low detuning (point A) where inelastic interdot tunnelling dominates. This field dependence vanishes at higher detuning where thermally activated tunnelling to the leads dominates.

As in previous work[4,11], we model spin evolution in (1,1) by treating the ensemble of nuclear spins within each dot as a static effective field $\mathbf{B}_{nuc}$ with slow internal dynamics, that adds to any applied Zeeman field (see Fig. 1b). $\mathbf{B}_{nuc}$ is randomly oriented with rms strength $B_{nuc} = b_0\sqrt{I_0(I_0+1)/N_{nuc}}$, where $b_0 = 3.5$ T is the hyperfine constant in GaAs, $I_0 = 3/2$ is the nuclear spin and $N_{nuc}$ is the effective number of nuclei with which the electron interacts[2,3,27,28]. In our dots, $N_{nuc}$ is estimated at $10^6 - 10^7$, giving $B_{nuc} \sim 2 - 6$ mT. The spins precess in a characteristic time $t_{nuc} = \hbar/g^*\mu_B B_{nuc} \sim 3 - 10$ ns, which can be regarded as an inhomogeneous dephasing time $T_2^*$. At $B = 0$, all four (1,1) spin states mix in this time, and tunnelling will appear insensitive to spin. With $B > B_{nuc}$ however, only (1,1)$T_0$ and (1,1)S are degenerate. These will continue to mix with the same rate, but (1,1)$T_+$ and $T_-$ will be frozen out.

To model this mixing, we assume static nuclear fields during each pulse, a spin-preserving inelastic interdot tunnelling rate $\Gamma_{in}$ from (1,1)S to (0,2)S, and a spin-independent rate $\Gamma_T$ due to thermally activated tunnelling via the (0,1) and (1,2) charge states. Zeeman eigenstates for two spins in fields $B\hat{\mathbf{z}} + \mathbf{B}_{nuc,l}$ and $B\hat{\mathbf{z}} + \mathbf{B}_{nuc,r}$, denoted $|(1,1)s,s'\rangle$ ($s,s'=\pm 1/2$), decay to (0,2)S based on their overlap with (1,1)S, with rates

$\Gamma_{s,s'} = \Gamma_{in} |\langle (1,1)S | (1,1)s,s' \rangle|^2$ as long as $\Gamma_{in} \ll g^*\mu_B B_{nuc}$. Averaging over nuclear field configuration and short-time dynamics gives decay rates for the $T_\pm$-like states

$$\Gamma_{\pm 1/2, \pm 1/2} = \frac{\Gamma_{in}}{4(1+(B/B_{nuc})^2)} \quad (1)$$

and $\Gamma_{\pm 1/2, \mp 1/2} = \Gamma_{in}/2 - \Gamma_{\pm 1/2, \pm 1/2}$ for the S-like and $T_0$-like states. At $B = 0$, total transition rates for all (1,1) states into (0,2)S are the same, $\tau_0^{-1} = \Gamma_{in}/4 + \Gamma_T$. For $B > B_{nuc}$, transitions rates $\tau_B^{-1} = \Gamma_{\pm 1/2, \pm 1/2} + \Gamma_T$ from $(1,1)T_\pm$ to (0,2)S are suppressed by field, while transitions from (1,1)S and $(1,1)T_0$ to (0,2)S are accelerated by up to a factor of two because they no longer mix with $(1,1)T_\pm$. During the gate-pulse transition from R to M, the relatively fast transition from (1,1)S to (0,2)S allows a fraction $q$ of the (1,1)S state to transfer adiabatically to (0,2)S, reducing the initial occupation of (1,1)S. The resulting average occupancy $N$ of (1,1) after a time $t_M$ is

$$N(t_M) = \frac{1}{t_M} \int_0^{t_M} dt \left( \frac{1}{2} e^{-t/\tau_B} + \frac{2-q}{4} e^{-t(2\tau_0^{-1} - \tau_B^{-1})} \right) \quad (2)$$

Experimentally measured values for $N$ as functions of $t_M$ and $B$ for various detunings are shown in Fig. 3, along with fits to the above theory. An additional field-independent parameter, $N_\infty$, accounts for nonzero $N(t_M)$ at long times due to thermal occupation of (1,1). $N_\infty = 0$ at large detuning but increases, as expected, near zero detuning. Nonzero $q$ values are found only at very low and very high detuning (where the R point is near zero detuning), where the slew rate of the pulse is low as it crosses to (0,2). With these parameters and $\tau_0$ set for a given detuning by fitting the zero field data (red), the high-field data (blue) are fit with just the longer decay times $\tau_B$ for the $(1,1)T_\pm$ states. The field-dependence curves (black) are then fully determined by $B_{nuc}$, which is most accurately determined from data in Fig. 4, as discussed below. Drift in sensor conductance over long field sweeps is compensated by allowing a vertical shift in the field-dependence curves. The depth and width of the dips in these curves is not adjustable.

Figure 4 shows the extracted decay times $\tau_0$ and $\tau_B$ versus detuning for various fields. As field increases, more points at high detuning fall along a line in this semi-log plot, denoting exponential energy dependence as expected for a thermally activated process. This persists over three orders of magnitude at the highest field, and with calibration from transport measurements yields a temperature of 160±20 mK. At zero field, thermal decay dominates only at the highest detunings, and the low-detuning times are well fit by a power law function of detuning with exponent 1.2±0.2 and offset 700 ns, typical of inelastic tunnelling in double quantum dots[29]. Adding these two processes gives the red curve in Fig. 4, in good agreement with the zero-field data. The 10 mT curve is fit using these zero-field parameters, but with times for the inelastic component increased by the factor $(1 + (B/B_{nuc})^2)$ from equation (1). The fit gives $B_{nuc}$=2.8±0.2 mT, or $N_{nuc} \sim 6 \times 10^6$, within expectations. This value uniquely determines the remaining theory curves. For $\tau_B$ longer than about 1 ms the decay is faster than theory predicts (though still $10^3$ times slower than at $B$=0), indicating that another mechanism such as spin-orbit coupling may operate on millisecond time scales[8-10]. Spin-orbit coupling is expected to dominate spin relaxation at external fields of several tesla[9]. This regime is better suited to parallel fields, which couple almost exclusively to spin, than to the perpendicular orientation used here, which affects orbital wavefunctions at high fields.

Given $B_{nuc}$ above, the model predicts an inhomogeneous dephasing time $T_2^* \sim 9$ ns for this device, which is independent of external field despite the enhanced relaxation times measured at higher fields. Up to 1 ms, the excellent agreement between experiment and theory indicates that hyperfine interaction is the only relevant source of spin relaxation in this system. Several strategies are available to circumvent this short dephasing time. Materials with zero nuclear spin, such as carbon nanotubes, avoid hyperfine effects entirely. Controlling **B**$_{nuc}$ via nuclear polarization[11,17] is tempting, however high polarization is required for $T_2$* to increase substantially[30]. An alternative is to use spin echo techniques such as pulsed ESR to extend coherence to the nuclear spin correlation time, expected to be of order 100 μs in these devices[4].

**Supplementary Information** accompanies the paper on *Nature*'s website (http://www.nature.com).

**Acknowledgements** We thank H.A. Engel and Peter Zoller for useful discussions. This work was supported by the ARO, the DARPA-QuIST programme, and the NSF including the Harvard NSEC.

**Competing Interests** The authors declare that they have no competing financial interests.

**Correspondence** and requests for materials should be addressed to C.M. (e-mail: marcus@harvard.edu).

**Figure 1** Measuring spin-selective tunnelling in a double quantum dot. **a**, Scanning electron micrograph of a device similar to the one measured. Metallic gates deplete a two-dimensional electron gas 100 nm below the surface, with density $2\times10^{11}$ cm$^{-2}$. A double dot is defined between gates L and R. Electrons can tunnel weakly between the dots and to conducting leads. The conductances $g_{ls}$ and $g_{rs}$ of the left and right quantum point contacts (QPCs) reflect the average occupation of each dot. **b**, Schematic of the (1,1) charge configuration. The spatially separated electrons feel different effective fields due to hyperfine interaction with the local *Ga* and *As* nuclei, in addition to a uniform externally applied field. **c**, Voltage pulses applied to gates L and R cycle through three configurations, Empty (E), Reset (R) and Measure (M). **d**, Right sensor conductance $g_{rs}$ as a function of dc voltages on the same two gates around the (1,1) to (0,2) transition, with cyclical displacements by the pulses shown by points E, R, and M. The dashed lines outline sensor plateaus indicating the stable charge configurations (0,1), (1,1), (0,2), and (1,2) during the M step. Inside the solid white-outlined "pulse triangle," the ground state is (0,2), but higher sensor conductance indicates that tunnelling is partially spin blocked. A plane is subtracted from the raw data to remove direct gate-QPC coupling. **e**, Energetics of the pulse sequence. In (0,2), only the singlet (0,2)S is accessible, whereas in (1,1), singlet and triplet are degenerate. (0,1) and (1,2) are both spin-1/2 doublets. Step E empties the second electron, leaving one electron in the right dot. Step R loads an electron into the left dot, occupying all four (1,1) states with equal probability. At M, (0,2)S is the ground state, but as tunnelling preserves spin, only (1,1)S and the $m_s$=0 triplet (1,1)T$_0$, with which it quickly mixes, can tunnel. Mixing of (1,1)T$_+$ and (1,1)T$_-$ with the singlet is weak away from zero field, hence their tunnelling is blocked. **f**, Energetics outside the pulse triangle. Below the pulse triangle, labelled M′, the one-electron state (0,1) has lower energy than (1,1) and provides an alternate, spin-independent path to (0,2). Above the pulse triangle, labelled M″, the three-electron state (1,2) provides an alternate spin-independent path.

**Figure 2** Dependence of the occupancy of the (1,1) state on measurement time, $t_M$, and external field, B. **a**, Charge sensor conductance, $g_{rs}$, as a function of $V_L$ and $V_R$ with short pulses ($t_M$ = 8 μs) at B=100 mT. Large average occupation of (1,1) is seen throughout the pulse triangle. Near the triangle edges, thermally activated tunnelling to the leads allows fast relaxation to (0,2), (see Fig. 1f). **b**, For longer pulses ($t_M$ = 80 μs), thermally relaxed triangle edges expand toward the centre of the triangle. **c**, At B=0, the (1,1) occupation is extinguished at low-detuning (near the (1,1)-(0,2) charge transition) as tunnelling to (0,2) becomes possible from the (1,1)T$_+$ and (1,1)T$_-$ states. **d**, Combine these two effects at zero field with long pulses, and no residual (1,1) occupation is seen, indicating complete relaxation to (0,2).

**Figure 3** Detailed measurements of blockaded (1,1) occupation. Average occupation <N> of the (1,1) charge state, based on calibrated charge sensor

conductances, at four detuning points (labelled A, B, C, D on the upper graph). Left panels show <N> as a function of $t_M$ at $B$ = 0 and $B$ = 150 mT. Middle panels show <N> as a function of $B$ for different $t_m$ times. Diagrams at right show schematically the relative position of energy levels and the extracted ratios of inelastic ($\Gamma_{in}$) to thermal ($\Gamma_T$) decay rates.

**Figure 4** Decay of (1,1) occupancy as a function of detuning at various magnetic fields. Dotted lines mark the locations of points A through D from Fig. 3. Fit of zero field theory (red curve) to data (red circles) sets all fit parameters except $B_{nuc}$, which is determined by fitting to the 10mT data (gold). Theory curves at other fields are then fully determined. Error bars at zero field result from the least squares fitting. Error bars at nonzero field reflect changes in the resulting decay rate as the zero-field fit parameters are varied within their uncertainties.

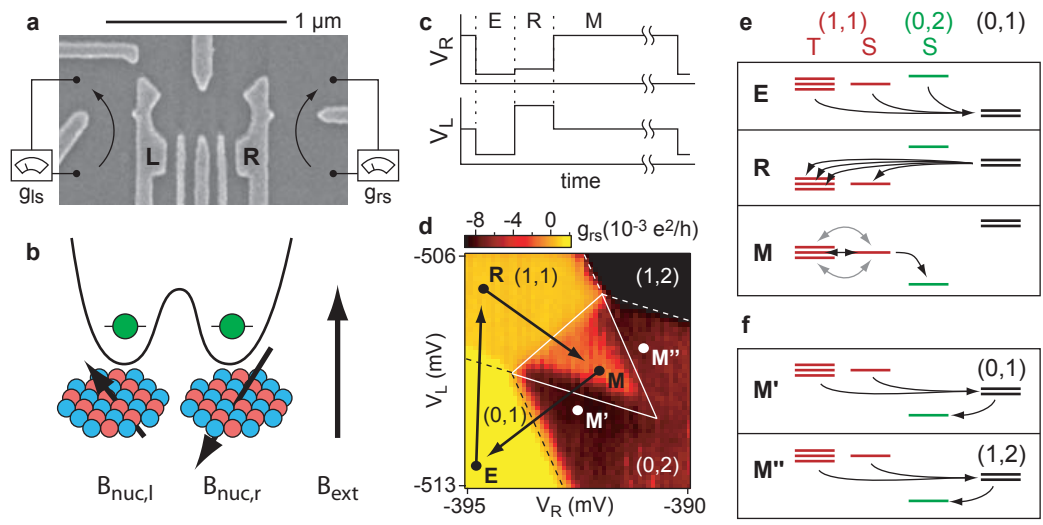

Figure 1

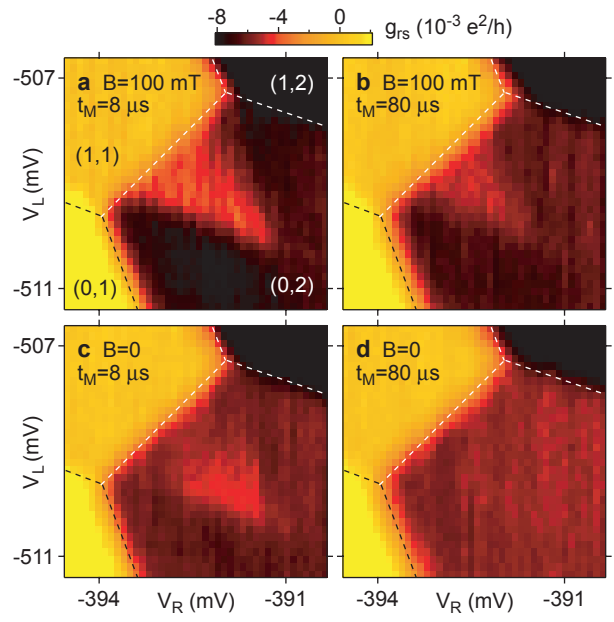

Figure 2

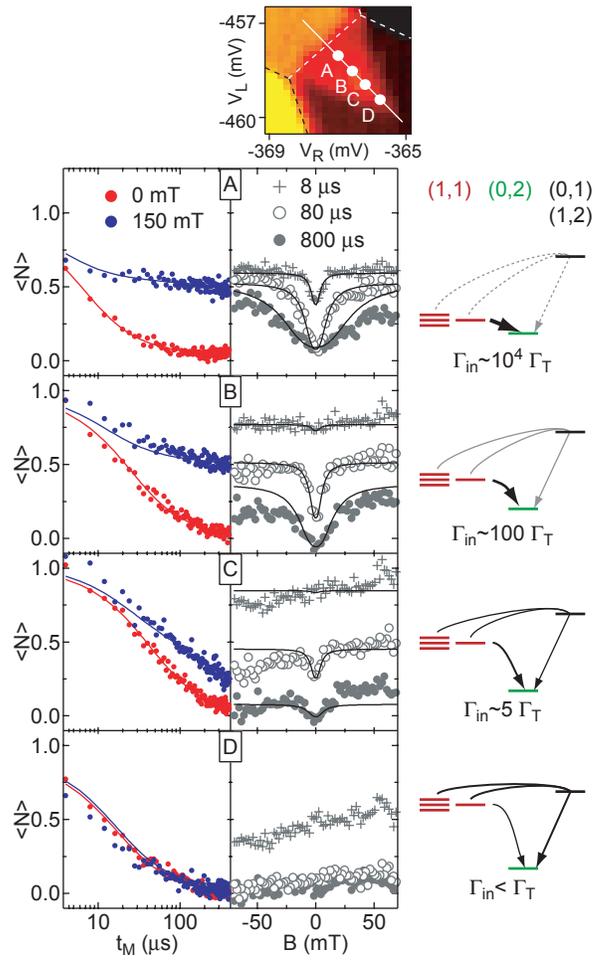

Figure 3

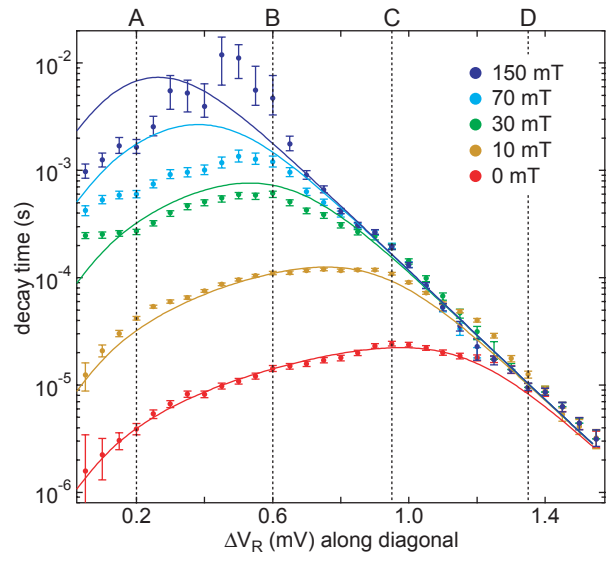

Figure 4

# Triplet-singlet spin relaxation via nuclei in a double quantum dot: supplementary information

A. C. Johnson[1], J. R. Petta[1], J. M. Taylor[1], A. Yacoby[1,2],
M. D. Lukin[1], C. M. Marcus[1], M. P. Hanson[3], A. C. Gossard[3]
[1]*Department of Physics, Harvard University,
Cambridge, Massachusetts 02138, USA*
[2]*Department of Condensed Matter Physics,
Weizmann Institute of Science, Rehovot 76100, Israel*
[3]*Materials Department, University of California,
Santa Barbara, California 93106, USA*

## I. EXPERIMENTAL SETUP

The experiment is implemented using a double quantum dot with local quantum point contact (QPC) charge sensors (Fig. 1a). These features are formed by metal gates on a GaAs/AlGaAs heterostructure containing a two-dimensional electron gas (2DEG). Voltage pulses are applied to gates L and R using a Tektronix AWG520 arbitrary waveform generator, with a rise time of ≤1 ns. The QPC conductances, $g_{ls}$ and $g_{rs}$, are measured using standard lock-in detection with 1 nA current excitations. $g_{ls}$ and $g_{rs}$ are sensitive to the local charge configuration and exhibit several percent changes when an electron either enters the double dot or moves from one dot to the other. We determine the absolute number of electrons in each quantum dot using charge sensing and tune the device to a region where the left dot contains 0 or 1 electron, while the right dot contains 1 or 2. The device is cooled in a dilution refrigerator to a temperature of about 100 mK and a magnetic field $\mathbf{B}_{ext} = B\mathbf{e}_z$ is applied perpendicular (along the $z$-axis) to the sample plane.

## II. DEFINITION OF $B_{\text{nuc}}$

We start by reviewing the effective magnetic field picture for nuclear spins in a single quantum dot[1], then extend it to the double-dot case. For a single electron in a single quantum dot with large orbital level spacing, $\hbar\omega \gg |g^*\mu_B B|$, we can write a spin hamiltonian for the ground orbital state of the dot, $\psi(\mathbf{r})$, and neglect higher orbital states. The included terms are the Zeeman interaction, $H_B = -g^*\mu_B \mathbf{B} \cdot \hat{\mathbf{S}}$, where $g^* = -0.44$ for GaAs and $\mu_B$ is the Bohr magneton, and the hyperfine contact interaction, $H_{HF} = Av_0 \sum_k |\psi(\mathbf{r}_k)|^2 \hat{\mathbf{I}}^k \cdot \hat{\mathbf{S}}$, where $A$ is the hyperfine constant for GaAs, $v_0$ the unit cell volume, and the sum is over all $N$ lattice sites the electron wavefunction overlaps with. We can define an effective (Overhauser) magnetic field due to nuclei,

$$\mathbf{B}_{\text{nuc}} = \frac{Av_0}{-g^*\mu_B} \sum_k |\psi(\mathbf{r}_k)|^2 \hat{\mathbf{I}}^k ,\qquad(1)$$

such that $H_{\text{tot}} = H_B + H_{HF} = -g^*\mu_B(\mathbf{B} + \mathbf{B}_{\text{nuc}}) \cdot \hat{\mathbf{S}}$ is the effective spin hamiltonian for a single electron quantum dot ($b_0 = \frac{A}{-g^*\mu_B} \sim 3.5$ tesla for GaAs[2]). At high temperatures relative to the nuclear Zeeman scale, $T \gg \hbar\gamma_n|\mathbf{B}|/k_B$, where $\gamma_n$ is the gyromagnetic ratio for a given nuclear spin species, the equilibrium state of each nuclear spin is well described by an identity density matrix, $\rho_k = \mathbf{1}/(2I_0+1)$. For $N \gg 1$, the equilibrium expectation values for $\mathbf{B}_{\text{nuc}}$ are gaussian distributed with zero mean and a root-mean-square magnitude[1,3]

$$B_{\text{nuc}} = \sqrt{\langle|\mathbf{B}_{\text{nuc}}|^2\rangle} = b_0[v_0^2 \sum_{kk'} |\psi(\mathbf{r}_k)|^2 |\psi(\mathbf{r}_{k'})|^2 \langle \hat{\mathbf{I}}^k \cdot \hat{\mathbf{I}}^{k'} \rangle]^{1/2}$$

$$= b_0\sqrt{v_0 I_0(I_0+1) \int d^3r \ |\psi(\mathbf{r})|^4} \ . \quad (2)$$

As the timescale for evolution of the nuclear field (set by nuclear dipole-dipole and back-action of the electron spin on the individual nuclear spins) is much longer than the precession time of the electron in the nuclear field, $\hbar/(g^*\mu_B B_{\text{nuc}})$, we assume the nuclear field to be static during the relevant electron-nuclear interaction time[1,3]. This assumption is consistent with the kHz linewidths for solid-state NMR on GaAs[2]. Using for $\psi(\mathbf{r})$ in Eq. (2) the ground state wave function for a parabolically confined quantum dot[4],

$$\psi(\mathbf{r}) = [\frac{16}{l^{3/2}} z \ e^{-4z/l}][\frac{\exp(-\frac{(x^2+y^2)}{4\sigma^2})}{\sqrt{2\pi\sigma^2}}], \quad (3)$$

with lateral dimension $\sigma$ and thickness $l$, yields an effective nuclear field $B_{\text{nuc}} = b_0\sqrt{I_0(I_0+1)}\sqrt{\frac{3v_0}{8\pi l\sigma^2}} = 2.3$ tesla $\sqrt{\frac{v_0}{l\sigma^2}} = 18$ mT $\sqrt{\frac{\hbar\omega[\text{meV}]}{l[\text{nm}]}}$, where $\hbar\omega$ is the single-particle level spacing of the dot. Comparing to a uniform wave function of volume $V$ ($\psi(\mathbf{r}) = V^{-1/2}$) gives an effective number of nuclear spins, $N = [v_0 \int d^3r \ |\psi(\mathbf{r})|^4]^{-1} = \frac{8\pi l\sigma^2}{3v_0}$ such that $\sqrt{\langle|\mathbf{B}_{\text{nuc}}|^2\rangle} = \frac{b_0\sqrt{I_0(I_0+1)}}{\sqrt{N}}$.

The spin hamiltonian for a double dot with one electron in each dot and negligible exchange coupling ($J = 0$) is given by

$$H_{(1,1)} = -g^*\mu_B[\mathbf{B}^{(1)} \cdot \hat{\mathbf{S}}^{(1)} + \mathbf{B}^{(2)} \cdot \hat{\mathbf{S}}^{(2)}] \ , \quad (4)$$

where $\mathbf{B}^{(i)} = \mathbf{B}_{\text{ext}} + \mathbf{B}_{\text{nuc}}^{(i)}$ is the total effective magnetic field for dot $i$. The eigenstates have spins aligned and anti-aligned with these two fields, $|s, s'\rangle$, with eigenenergies $E_{ss'} = -g^*\mu_B(B^{(1)}s + B^{(2)}s')$ and $s, s' = \pm\frac{1}{2}$. Eigenstates of external field (when $B_{\text{nuc}} = 0$) are denoted $|\tilde{s}, \tilde{s}'\rangle$, and are spin aligned/anti-aligned with the total external magnetic field. Without loss of generality we choose the external field to lie along the $z$ axis.

### III. FITTING FUNCTION

#### A. Hyperfine-driven decay

The two-step relaxation process we consider—spin mixing due to nuclei followed by inelastic decay—closely resembles processes well known in the chemical physics literature[3] but not previously investigated in the context of quantum dots[5-8] to our knowledge. In this model, energy relaxation is mediated by a process that couples only to charge, leaving the spin state unchanged. In the absence of spin selection rules, this would couple any (1,1) state to any (0,2) state with a transition rate $\Gamma_{in}$ that depends on interdot coupling as well as the energy difference between the two charge states ($E_{SG}$). Such inelastic decay, for example due to phonons, has been studied previously in double-dot systems[9]. For the spin-mixing part, we take $\hbar\Gamma_{in}, J \ll |E_{ss'}|$ and work in the eigenbasis $|s, s'\rangle$ set by local nuclear fields. The decay rate of $|s, s'\rangle$ into the (0,2) singlet, $|G\rangle$, is given by $\Gamma_{ss'} = \Gamma_{in}|\langle s, s'|S\rangle|^2$, i.e., the overlap of $|s, s'\rangle$ with the (1,1) singlet, $|S\rangle$, times the rate of decay of $|S\rangle$ into $|G\rangle$. We remark that this model bears a strong resemblance to the model for a single quantum dot with two electrons of Ref. 5.

We now evaluate the overlap matrix elements, $|\langle s, s'|S\rangle|^2$. A single spin state $|s\rangle$ for a magnetic field $\mathbf{B} = (x, y, z)$ (and of norm $n = \sqrt{x^2 + y^2 + z^2}$) can be expressed in the basis of eigenstates of external field ($|\tilde{s}\rangle$, spin aligned with the $z$-axis) as

$$|s\rangle = \frac{-(2s)i(n+z)|\tilde{s} = s\rangle - (ix - 2sy)|\tilde{s} = -s\rangle}{\sqrt{2n^2 + 2nz}} \ . \tag{5}$$

where $s$ and $\tilde{s}$ can take the values $\pm 1/2$. In the tilde-basis, $|S\rangle = (\left|\frac{\tilde{1}}{2}, -\frac{\tilde{1}}{2}\right\rangle - \left|-\frac{\tilde{1}}{2}, \frac{\tilde{1}}{2}\right\rangle)/\sqrt{2}$. The overlap elements, given a field $B_1 = (x_1, y_1, z_1)$ in dot 1 and $B_2 = (x_2, y_2, z_2)$ in dot 2, are then

$$\left|\left\langle S\left|\pm\frac{1}{2}, \pm\frac{1}{2}\right.\right\rangle\right|^2 = \frac{[(n_1 + z_1)x_2 + x_1(n_2 + z_2)]^2 + [(n_1 + z_1)y_2 + y_1(n_2 + z_2)]^2}{8(n_1^2 + n_1 z_1)(n_2^2 + n_2 z_2)} \ , \tag{6}$$

$$\left|\left\langle S\left|\pm\frac{1}{2}, \mp\frac{1}{2}\right.\right\rangle\right|^2 = \frac{[(n_1 + z_1)(n_2 + z_2) + x_1 x_2 + y_1 y_2]^2 + [-x_1 y_2 + y_1 x_2]^2}{8(n_1^2 + n_1 z_1)(n_2^2 + n_2 z_2)} \ . \tag{7}$$

The states with the same $|m_s|$ value have the same overlap. Averaging over all quasi-static field values, we can find $\langle\Gamma_{ss'}\rangle$, the effective decay rate for an experiment with many different realizations, each with a different field value drawn from the field distribution. As we have taken the external field to be aligned along the $z$-axis,

$$\langle z_i \rangle = B \ ,$$
$$\langle x_i \rangle = \langle y_i \rangle = 0 \ ,$$
$$\langle (\mu_i - \delta_{\mu z} B)(\nu_j - \delta_{\nu z} B) \rangle = \delta_{ij} \delta_{\mu\nu} B_{\text{nuc}}^2 / 3 \ .$$

We assume, appropriate to the high temperature approximations used heretofore, that the nuclear fields in dots 1 and 2 are uncorrelated; accordingly, expectation values of two separate fields factorize. This yields

$$\langle \Gamma_{\pm\frac{1}{2}, \pm\frac{1}{2}} \rangle = \Gamma_{in} \langle F \rangle \langle G \rangle \tag{8}$$

$$\langle \Gamma_{\pm\frac{1}{2}, \mp\frac{1}{2}} \rangle = \frac{\Gamma_{in}}{2}(\langle F \rangle^2 + \langle G \rangle^2) \tag{9}$$

where

$$\langle F \rangle = \left\langle \frac{(n+z)^2}{2n^2 + 2nz} \right\rangle = \left\langle \frac{n+z}{2n} \right\rangle$$

$$\langle G \rangle = \left\langle \frac{x^2 + y^2}{2n^2 + 2nz} \right\rangle = \left\langle \frac{n^2 - z^2}{2n(n+z)} \right\rangle = \left\langle \frac{n-z}{2n} \right\rangle$$

are averages over a single dot. We have assumed both dots to be of equal size, having the same effective strength on the average. Transforming to cylindrical coordinates, the integral over a gaussian corresponding to the single-dot nuclear field distribution is

$$I = \left\langle \frac{z}{n} \right\rangle = \frac{1}{[2\pi B_{\text{nuc}}^2/3]^{3/2}} \int_{-\infty}^{\infty} dz \int_0^{\infty} r \, dr \int_0^{2\pi} d\theta \, \frac{z \, \exp[-\frac{r^2 + (z-B)^2}{2B_{\text{nuc}}^2/3}]}{\sqrt{r^2 + z^2}} \ . \tag{10}$$

After integration over $\theta, r$ and variable change, $u = \sqrt{3/2} \frac{z}{B_{\text{nuc}}}$,

$$I = e^{-3B^2/2B_{\text{nuc}}^2} \int_{-\infty}^{\infty} du \, u \, e^{\sqrt{6}uB/B_{\text{nuc}}} \text{Erfc}(|u|) \ . \tag{11}$$

We note that to a good approximation, $I \simeq \frac{\langle z \rangle}{\sqrt{\langle n^2 \rangle}} = \frac{|B|}{\sqrt{B_{\text{nuc}}^2 + B^2}}$ for all external field values, $B$. This approximation intuitively corresponds to Zeeman-split levels broadened by the nuclear field, and breaks down for larger $\Gamma_{in}$ or $J$ values. Using this approximation, we find the effective decay rates, used in the main text, from the four eigenstates of the (1,1) charge configuration to the (0,2) singlet state to have two forms, one for the $|m_s| = 1$ states $(T_+, T_-)$, with a rate $\Gamma_{\pm\frac{1}{2},\pm\frac{1}{2}} = \frac{\Gamma_{in}}{4} \frac{B_{\text{nuc}}^2}{B_{\text{nuc}}^2 + B^2}$, and the other for the $|m_s| = 0$ states $(T_0, S)$, with a rate $\Gamma_{\pm\frac{1}{2},\mp\frac{1}{2}} = \frac{\Gamma_{in}}{2}[1 - \frac{B_{\text{nuc}}^2}{2(B_{\text{nuc}}^2 + B^2)}]$.

### B. Thermal component

In addition to the inelastic decay from the (1,1) singlet to the (0,2) singlet, coupling to the leads allows for a spin-independent transition of either $(1,1) \to (1,2) \to (0,2)$ or $(1,1) \to (0,1) \to (0,2)$ to occur, breaking blockade and reducing the expected signal. As a thermally activated process, the rate for each should dependend on the energy difference between the (1,1) state and either (1,2) or (0,1). Denoting this detuning $E_T$, the corresponding decay has the expected form, $\Gamma_T = \Gamma_0 e^{-E_T/k_B T}$. We note that larger (1,1) to (0,2) detuning $E_{SG}$ corresponds to smaller $E_T$ detuning, as the M point moves closer to the top of the triangle of Fig. 2. This functional form is consistent with the observed high detuning behavior shown in Fig. 4. Combining this decay with the previous section results, it is convenient to define $\tau_0^{-1} = \Gamma_{in}/4 + \Gamma_T$, the zero-field decay time, and $\tau_B^{-1} = \Gamma_{\pm\frac{1}{2},\pm\frac{1}{2}} + \Gamma_T$, the decay rate of the $|m_s| = 1$ states.

---